# Interferometric evidence of non-volatile anomalous phase shifts in exchange-spin-split Josephson supercurrent diodes

Kun-Rok Jeon,[1*†] Jae-Keun Kim,[2†] Jiho Yoon,[2] Jae-Chun Jeon,[2] Hyeon Han,[2,3]

Audrey Cottet,[4,5] Takis Kontos,[4,5] and Stuart S. P. Parkin[2*]

[1]*Department of Physics, Chung-Ang University (CAU), 06974 Seoul, Republic of Korea*

[2]*Max Planck Institute of Microstructure Physics, Weinberg 2, 06120 Halle (Saale), Germany*

[3]*Department of Materials Science and Engineering, Pohang University of Science and Technology (POSTECH), 37673 Pohang, Republic of Korea*

[4]*Laboratoire de Physique de l'Ecole Normale Supérieure, ENS, Université PSL, CNRS, Sorbonne Université, Université Paris-Diderot, Sorbonne Paris Cité, 75005 Paris, France.*

[5]*Laboratoire de Physique et d'Etude des Matériaux, ESPCI Paris, PSL University, CNRS, Sorbonne Université, 75005 Paris, France.*

*Corresponding author. Email: jeonkunrok@gmail.com (K.-R.J.); stuart.parkin@halle-mpi.mpg.de (S.S.P.P.)

†These authors contributed equally to this work.

## ABSTRACT

**The recent realization of zero-field, polarity-reversible supercurrent rectification in proximity-magnetized Rashba-type Pt Josephson junctions (JJs) enables the development of superconducting logic circuits and cryogenic memory applications. Here, we demonstrate a *non-volatile* anomalous phase shift $\varphi_0$ directly probed via superconducting quantum interferometry, providing phase-sensitive evidence of spontaneous time-reversal symmetry breaking in these Rashba-type systems. By replacing the Pt barrier with 5d or 4d element layers exhibiting different (para-)magnetic susceptibilities, spin-orbit coupling strengths, and electronic band structures, we elucidate the role of proximity effects in governing zero-field diode behavior. Ta (W) JJs exhibit zero-field diode efficiencies of ~17% (~5%) at 2 K, which are slightly (significantly)**

**lower than those of Pt JJs. Notably, the diode polarity in Ta and W JJs is reversed relative to Pt JJs. Combined with the large zero-field diode efficiency (~15% at 2 K) observed in highly magnetic-susceptible Pd JJs, these results show that non-volatile $\varphi_0$ and, consequently, zero-field diode performance can be tuned through proximity engineering of interfacial magnetic ordering and Rashba spin-orbit interaction.**

Spontaneous breaking of time-reversal symmetry in a non-centrosymmetric conductor leads to a non-reciprocal electronic transport (*1-4*) at zero external magnetic field. This so-called zero-field and polarity-switchable rectification has been recently observed in the supercurrents of proximity-magnetized Rashba(-type) Pt Josephson junctions (JJs) (*5*), Co- or Fe-inserted non-centrosymmetric Nb/V/Ta artificial superlattices (*6,7*), gated small-twist-angle trilayer graphene (*8*), cross-like Nb planar JJs with an artificial vortex trap (*9*), and Nb/EuS proximity bilayers (*10*). The experimental realization of such zero-field polarity-switchable supercurrent rectification especially in the form of lateral JJs (*5*) provides a technical progress for the development of ultra-low-power rectifier circuits and non-volatile (superconducting-)phase memories (*11-13*).

Zeeman-spin-splitting-driven Josephson supercurrent diodes (*14-22*) with a non-magnetic inversion-asymmetric barrier necessarily require an external magnetic field $\mu_0 H$ to break the time-reversal symmetry. This applied $\mu_0 H$ induces the non-zero ground-state anomalous phase-shift $\varphi_0 \neq 0$ (*20-22*) in the current-phase relationship (CPR) (*23*), resulting in a non-sinusoidal CPR characterized by $I_s(\varphi) \neq -I_s(-\varphi)$, where $\varphi$ is the Josephson phase.

This non-sinusoidal CPR, in conjunction with higher-order harmonics, leads to Josephson supercurrent non-reciprocity (*12-22*), manifesting as unequal positive and negative critical currents $I_c^+ \neq |I_c^-|$. Here $\varphi_0$ scales with spin-splitting fields, the degree of inversion asymmetry and spin-orbit coupling (SOC) (*17-22*), and the underlying CPR is approximately described as $I_s(\varphi) \approx I_{c1} sin(\varphi + \varphi_0) + I_{c2} sin(2\varphi)$. For a small second-harmonic contribution and small $\varphi_0$, $Q = \frac{I_c^+ - |I_c^-|}{I_c^+ + |I_c^-|} \approx -\frac{I_{c2}\varphi_0}{I_{c1}}$ (see Supplementary Material for details). Note that the combination of $\varphi_0$ and $I_{c2} sin(2\varphi)$ leads to an anisotropic $\mu_0 H$-dependent $I_c$ (*20-23*) for the Zeeman-spin-splitting-driven Josephson supercurrent diodes (*14-22*). On the other hand, exchange-spin-splitting-driven Josephson diodes (*24-26*) with a magnetic inversion-asymmetric barrier, like the proximity-magnetized Pt JJs (*5*) with Rashba SOC (*27*), can inherently possess $\varphi_0$ even in the absence of $\mu_0 H$. This results in the non-volatility of $\varphi_0$ for $\mu_0 H = 0$, which, in turn, enables the spontaneous magnetization of $\varphi_0$-barrier to induce the zero-field anisotropic $I_c$ (*5,24-26*) and to control the polarization of supercurrent non-reciprocity by preconfiguring its remanent state. These exchange-spin-splitting-driven Josephson diodes (*5,24-26*) are thus highly promising for future superconducting-phase memory and logic circuit applications (*11-13*). For instance, instead of using a $\varphi$ $(= 0$ or $\pi)$ junction (*13*) as a memory cell for the Josephson phase memory, where the necessary arbitrary phase shift $\varphi$ in the ground state is obtained by geometrically combining 0 and $\pi$ JJs in parallel, the exchange-spin-split Josephson $\varphi_0$-junctions having $0 < \varphi_0 < \pi$ even in the form of a single junction can be utilized.

So far, unlike the case of the former Zeeman-spin-split Josephson $\varphi_0$-junctions (*17-20*), the interferometry characterization (*17-20*) of $\varphi_0$ at zero magnetic field in the latter exchange-spin-split Josephson $\varphi_0$-junctions and how this *non-volatile* $\varphi_0$ is linked to the zero-field non-reciprocal supercurrents remain to be investigated. In addition, a systematic study of how the choice of a proximity layer for the Josephson $\varphi_0$-barrier affects the sign and magnitude of the resulting supercurrent non-reciprocity is outstanding not only for disentangling the intrinsic origin of zero-field supercurrent non-reciprocity from other extrinsic factors (*e.g.* spatially non-uniform supercurrents and rectifying motion of Josephson or Abrikosov vortices) (*9,10*) but also for further improving/tuning the Josephson diode performance based on the more fundamental quantity of $\varphi_0$ rather than the Josephson current-voltage characteristics.

In this work, by substituting the Pt Josephson $\varphi_0$-barrier (*5*) for either 5d or 4d element proximity layer (Figs. 1A and S1) with different (para-)magnetic susceptibility, SOC and electronic band structure, and fabricating a superconducting quantum interference device (SQUID) consisting of two symmetric Josephson $\varphi_0$-junctions (Fig. 1B), we are able to elucidate the role of interface magnetic ordering and Rashba effect in determining the zero-field Josephson diode properties, and to prove the existence of the non-volatile $\varphi_0$ and its correlation with the zero-field diode effect. In our experimental setup, a pair of Josephson $\varphi_0$-junction and SQUID are fabricated on the same proximity layer (Fig. 1C), which is (proximity-)magnetized by a ferrimagnetic insulator $Y_3Fe_5O_{12}$ (YIG) underneath it, allowing for a direct comparison between the non-reciprocal Josephson supercurrents and phase-resolved interferometer data. Note also that our SQUID geometry is conceived to detect both non-volatile $\varphi_0$ and the zero-field diode effect even in a single SQUID (Fig. S2), as discussed below.

## RESULTS AND DISCUSSION

We first investigate how the zero-field diode properties rely on the choice of a proximity layer for the Josephson $\varphi_0$-barrier. Figure 2A exhibits *zero-field* current-voltage *I-V* curves of the magnetic 4-nm-thick Pt JJ, normalized by $I_c^{avg} = \frac{I_c^+ + |I_c^-|}{2}$, for two different magnetization orientations $M_{Pt}$ (// $\pm x$-axis) below the junction's superconducting transition temperature $T_c$, taken at the temperature $T$ = 2 K. As consistent with our recent experiment (*7*), visible zero-field supercurrent non-reciprocity $\Delta I_{c\,,\,\mu_0 H\,=\,0} = I^+_{c,\ \mu_0 H\,=\,0} - \left|I^-_{c,\,\mu_0 H\,=\,0}\right| \neq 0$ is detected and its polarity is clearly reversed ($\Delta I_{c,\,\mu_0 H\,=\,0} > 0 \rightarrow\ \Delta I_{c,\,\mu_0 H\,=\,0} < 0$) when $M_{\mathrm{Pt}}$ flips from the positive to the negative $x$-direction, further proving the zero-field, polarity-switchable Josephson supercurrent non-reciprocity in the exchange-spin-split Pt JJs (*5*). By replacing the Pt Josephson $\varphi_0$-barrier with a 5d heavy metal (HM = Ta or W) layer in Figs. 2B and 2C, we find two intriguing HM-dependent diode properties. First, the zero-field diode efficiency $\left|Q_{\mu_0 H\,=\,0}\right| = \left|\frac{\Delta I_{c,\,\mu_0 H\,=\,0}}{2 I^{avg}_{c,\ \mu_0 H\,=\,0}}\right| \approx 17\%$ obtained from the 4-nm-thick Ta JJ (Fig. 2B) is comparable to that of the 4-nm-thick Pt JJ (Fig. 2A) whereas the 4-nm-thick W JJ (Fig. 2C) reveals a much smaller $\left|Q_{\mu_0 H\,=\,0}\right| \approx 5\%$. Second, the zero-field diode polarity $Q_{\mu_0 H\,=\,0} > 0$ ($Q_{\mu_0 H\,=\,0} < 0$) of the Ta and W JJs appears *opposite* to that $Q_{\mu_0 H\,=\,0} < 0$ ($Q_{\mu_0 H\,=\,0} > 0$) of the Pt JJs for the positive (negative) $x$-direction remanent-state $M_{\mathrm{HM}}$.

To understand these, we have developed our own theoretical model, analyzing a metal with Rashba spin–orbit interaction connected to two SCs and a ferromagnetic insulator. Using the Usadel equations, we describe the propagation of both singlet and triplet superconducting correlations in the diffusive limit (see Supplementary Material for full details). From this analysis, we find that $Q_{\mu_0 H\,=\,0} \propto \varphi_0 = -\tilde{\alpha}\tilde{\gamma}_\phi L^2 = -\frac{\alpha\gamma_\phi}{\sqrt{8m^2D^3\hbar|\omega_n|}}L^2 \approx -\frac{\alpha\gamma_\phi}{\sqrt{8m^2D^3\hbar|\pi k_B T|}}L^2$ with $\gamma$ = 90° for a diffusive ($l_{mfp} < \xi,\ l_{mfp} < L$) junction. Here $\tilde{\alpha} = \frac{\alpha}{2mD}$ and $\tilde{\gamma}_\phi = \frac{\gamma_\phi}{\sqrt{2\hbar D|\omega_n|}}$. $\alpha$ is the Rashba coefficient, $m$ is the effective electron mass, $D$ is the diffusion coefficient, $\gamma_\phi = -\Delta E_{ex}$ represents the exchange spin-splitting, $\hbar$ is the reduced Planck constant, $\omega_n$ ($\approx \pi k_B T$ in the high $T$ limit) is the Matsubara frequency and $\gamma$ is the azimuthal angle (Fig. 1A). $l_{mfp}$, $\xi$ and $L$ are the mean free path, the coherence length and the spacing of the Josephson $\varphi_0$-barrier (Fig. 1A), respectively. Since the electrical resistivity of Pt, Ta, W, and Pd layers, which is related with $m$ and $D$, does not significantly differ among these materials (Fig. S2) and $L$ is fixed at $80 - 100$ nm (see the Methods), it is reasonable to speculate that $\gamma_\phi$ and $\alpha$ predominantly govern the polarity and amplitude of the Josephson diode efficiency over $m$ and $D$. Earlier theories (*28,29*), in fact, pointed out that the *magnitude and sign* of $\gamma_\phi$ are strongly dependent on the interface properties.

Based on the results of the anomalous Hall response (Fig. S3) and non-local thermal magnon transport (Fig. S4), which respectively characterize the magnetic proximity effect and the spin-to-charge conversion, we now explain the two aforementioned properties: the proximity-layer-dependent diode strength and polarity. The W layer exhibits a smaller anomalous Hall response in Fig. S3 (while showing a comparable spin-to-charge conversion efficiency in Fig. S4) compared to the Pt and Ta layers. Given this, the first property is explained in term of the weak proximity-induced $\Delta E_{ex}$ in the W layer, as in line with its very low (para-)magnetic susceptibility (*28*). This explanation can also be supported by a large $|Q_{\mu_0 H\,=\,0}| \approx$ 15% (Fig. 2D) attained from the JJ with a highly magnetic-susceptible Pd layer (*30,31*) (Fig. S3) having a modest SOC strength (Fig. S4). The second property relates to how the overall sign of $\gamma_\phi \alpha$ is given in our structure, in which the HM Josephson barrier is sandwiched between top $AlO_x$ and bottom YIG layers (see the Methods for details). The sign of $\gamma_\phi (= -\Delta E_{ex})$ relies on whether the magnetic proximity coupling (*32*) of the HM is interlayer-ferromagnetic or interlayer-antiferromagnetic with the YIG magnetization whereas the sign of $\alpha$ is determined by the sum of potential gradient (*27,32*) at the top $AlO_x$/HM interface and that at the bottom HM/YIG interface. This indicates that the proximity magnetic ordering across the bottom HM/YIG interface and the Rashba effect at the *top* $AlO_x$/HM interface are both responsible for the sign-reversed diode effect in the Ta and W JJs. In the framework of standard spin pumping theory, interfacial spin-transfer is typically described by a complex quantity known as the spin mixing conductance $G_{\uparrow\downarrow} = G_r + iG_i$, where the transfer of the interfacial exchange field is included in the imaginary part of the spin mixing conductance, $G_i$. We also note that the sign of $\gamma_\phi (= -\Delta E_{ex})$, which reflects whether the HM/YIG coupling is inter-layer ferromagnetic or antiferromagnetic, is theoretically linked to the sign of $G_i$. However, to the best of our knowledge, there is no direct experimental demonstration showing that a sign change in $G_i$ necessarily guarantees an inversion of $\gamma_\phi (= -\Delta E_{ex})$, leaving this to be firmly confirmed in future studies.

By plotting $Q_{\mu_0 H\,=\,0}$ as a function of $T/T_c$ for four different JJs (Fig. 2E), we witness the progressive enhancement of $Q_{\mu_0 H\,=\,0}$ with decreasing $T/T_c$ and for all the JJs, $Q_{\mu_0 H\,=\,0}(T/T_c)$ is well fitted by a $\sqrt{1-\frac{T}{T_c}}$ function. This proximity-layer-independent $T$-evolving behaviour provides an experimental signature of the exchange-spin-split Josephson $\varphi_0$-junctions (*24-26*). In fact, our theory predicts that as $T/T_c$ goes down, $\Delta E_{ex}$ and $\alpha$ in the numerator [ $D$ and $\omega_n$ ( $\approx \pi k_B T$ ) in the denominator] of $\varphi_0$ increase (decrease), which qualitatively explains the *effective* enhancement of $Q_{\mu_0 H\,=\,0} \propto \varphi_0 \propto \frac{\gamma_\phi \alpha}{D^{1.5}\sqrt{T}}$ (for $\gamma$ = 90°) at a lower $T/T_c$. Furthermore, the predicted and experimentally observed reduction of higher harmonics with increasing $T/T_c$ (*33*), which suppresses the nonreciprocity $Q \approx -\frac{I_{c2}\varphi_0}{I_{c1}}$ at higher $T/T_c$, is well consistent with our observations.

If the Josephson $\varphi_0$-barrier with Rashba SOC is non-magnetic (*14-22*), $Q$ scales linearly with the strength of in-plane (IP) magnetic field $\mu_0 H_\parallel$ in the low-field regime, *i.e.* magneto-linearity (*14,16,20*). On the other hand, if it is proximity-magnetized (*24-26*), $Q(\mu_0 H_\parallel)$ mimics the $\mu_0 H_\parallel$-driven reversal of the $\varphi_0$-barrier's magnetization $M_{\varphi_0-barrier}$, *i.e.* magneto-hysteresis (*5,7*). This hysteretic characteristic of exchange-spin-split $\varphi_0$-junctions can be seen in the $Q(\mu_0 H_\parallel)$ plots of our Pt, Ta, W and Pd JJs at a fixed $\gamma$ = 90° (Figs. 3A-3D). Here all the junctions exhibit the low-field magneto-hysteresis for $|\mu_0 H_\parallel| \leq 5$ mT, above which $|Q|$ diminishes monotonically, regardless of the diode polarity. We note that for Zeeman-spin-split $\varphi_0$-junctions (*17-20*), the magnetic field dependent $\Delta I_c(\mu_0 H_\parallel)$ is controlled by the two competing effects (*17*) of 1) the anomalous phase shift $\varphi_0(\mu_0 H_\parallel) \propto \mu_0 H_\parallel$ versus 2) the ratio of first and second harmonic terms $\frac{I_{c2}(\mu_0 H_\parallel)}{I_{c1}(\mu_0 H_\parallel)} \propto \frac{\Delta(\mu_0 H_\parallel)}{\Delta_0}$, where $\Delta(\mu_0 H_\parallel)$ is the suppressed superconducting energy gap in the presence of $\mu_0 H_\parallel$ and $\Delta_0$ is the zero-field superconducting gap *at the Josephson barrier*. Since in our exchange-spin-split $\varphi_0$-junctions, the internal exchange field dominates the necessary spin-splitting of the Josephson barrier (or, proximity layer) over the external magnetic field, one can assume that 1) $\varphi_0(\mu_0 H_\parallel)$ remains constant and consequently, conclude that 2) the pair breaking factor $\frac{I_{c2}(\mu_0 H_\parallel)}{I_{c1}(\mu_0 H_\parallel)} \propto \frac{\Delta(\mu_0 H_\parallel)}{\Delta_0}$ *at the Josephson barrier* is responsible for the monotonic decrease of $\Delta I_c(\mu_0 H_\parallel)$ observed in our $\varphi_0$-junctions (Fig. 3). A rather slow suppression of $|Q|$ for the W JJ in $|\mu_0 H_\parallel| > 5$ mT (Fig. 3C) is also understood in terms of a non-negligible relative contribution of Zeeman-spin-splitting to the *weak* exchange-spin-splitting induced in the W proximity layer (Fig. S3).

The magneto-chirality, $Q \propto \varphi_0 \times sin(\gamma)$ in the limit of a small second-harmonic contribution and small $\varphi_0$ (*5,7,14,16,20*), another characteristic of the Josephson $\varphi_0$-junctions (*14-22,24-26*) with isotropic SO fields (e.g. Rashba SOC), is also checked by measuring $Q$ as a function of $\gamma$ for our Pt, Ta, W and Pd JJs at a constant $\mu_0 H_\parallel$ = 5 mT. As summarized in Fig. 3E, for all the JJs, the $Q_{\mu_0 H_\parallel = 5\ mT}(\gamma)$ data are well fitted by a sine function, further supporting an intrinsic $\varphi_0$-origin (*14-22,24-26*) of their Josephson supercurrent non-reciprocity.

To directly probe the non-volatile $\varphi_0$ in our exchange-spin-split JJs and to correlate it with $Q_{\mu_0 H = 0}$, we next execute the phase-resolved interferometry characterization using a direct-current (d.c.) SQUID (*12,17-20*) in the voltage state (*20,34*). Our SQUID geometry is designed for supercurrents ($I_s^1$, $I_s^2$) of two symmetric Josephson $\varphi_0$-junctions to flow orthogonal to each other (Fig. 1C), enabling measurements of both non-volatile $\varphi_0$ and $Q_{\mu_0 H = 0}$ in a single SQUID (*12,20*) (Fig. S2). Note also that in our SQUID geometry, the non-zero $Q_{\mu_0 H = 0}$ can be detected by setting the directions of $I_s^1$ and $I_s^2$, respectively, perpendicular and parallel to the remanent-state $M_{\varphi_0 - barrier}$, or vice versa (Fig. S5). When the SQUID, which is composed of overdamped JJs with no hysteresis (Figs. 2A-2D) and in the limit of small self-inductance with negligible screening (see Supplementary Information for a quantitative discussion), is $I$-biased, the conversion of a magnetic flux into (time-averaged) $V$ modulation for the ±$x$-direction remanent-state $M_{\varphi_0-barrier}$ in the low-field limit is given by (*20,31*) $V_{\pm}(\mu_0 H_{\perp}, I) = \frac{R_n}{2}\sqrt{(I)^2 - \left(2I_c^{avg} cos\left(\pi \frac{\Phi_{SQUID}}{\Phi_0} \pm \frac{\varphi_{tot}}{2}\right)\right)^2}$. Here we assume that two symmetric JJs constitute the SQUID, $R_n$ ($I_c^{avg}$) is the normal-state zero-bias resistance (averaged critical current) of each JJ, $\Phi_{SQUID} = \mu_0 H_{\perp} A_{SQUID}^{eff}$ is the magnetic flux threading the SQUID loop given by $\mu_0 H_{\perp}$ and $A_{SQUID}^{eff} = L_x^{ctc} L_y^{ctc} \approx 11\ \mu\text{m}^2$ (Fig. 1B and 1C, Ref. *30*), and $\Phi_0 = \frac{h}{2e} = 2.07 \times 10^{-15}$ T·m$^2$ is the magnetic flux quantum. $\varphi_{tot} = \varphi_0^1 \sin\left(\frac{\pi}{4}\right) + \varphi_0^2 \sin\left(\frac{\pi}{4}\right) \approx 2\varphi_0 \sin\left(\frac{\pi}{4}\right)$, where the pre-factor 2 accounts for the doubled anomalous phases acquired by two symmetric JJs around the SQUID loop and the post-factor $\sin\left(\frac{\pi}{4}\right)$ takes the magneto-chiral effect into account (Fig. S1). As previously pointed out (*17-20*), it is challenging to accurately determine the absolute $\varphi_0$ of the constituent JJs due to a non-vanishing screening effect (*35*) in the SQUID loop under application of $\mu_0 H_{\perp}$. Therefore, we below focus our analysis on the horizontal difference of $V_+(\mu_0 H_{\perp}, I)$ versus $V_-(\mu_0 H_{\perp}, I)$, which provides a reliable measure of $\varphi_0 = \frac{\varphi_{tot}}{2\sin\left(\frac{\pi}{4}\right)}$.

Clearly, as presented in Figs. 4A-4H, $V_-(\mu_0 H_{\perp})$ horizontally shifts with respect to $V_+(\mu_0 H_{\perp})$ and the details depend on the type of the constituent JJs. Qualitatively, the relative horizontal shift is positive (negative) for the Pt JJ-based (Ta JJ-based) SQUID while it is almost unchanged for the W JJ-based SQUID. Moreover, the $V_{\pm}(\Phi_{SQUID}/\Phi_0)$ data of Pd JJ-based SQUID indicates the positive relative horizontal shift, which is the same as that of the Pt JJ-based SQUID. Best fits to the measured $V_{\pm}(\Phi_{SQUID}/\Phi_0)$ data (Figs. 4A, 4B, and 4C) using the above formula give $\varphi_{tot} = (+0.62 \pm 0.03)\pi$, $(-0.46 \pm 0.03)\pi$, $(-0.04 \pm 0.02)\pi$ and $(+0.41 \pm 0.01)\pi$, corresponding to $\varphi_0 = (+0.44 \pm 0.02)\pi$, $(-0.33 \pm 0.02)\pi$, $(-0.03 \pm 0.01)\pi$ and $(+0.29 \pm 0.01)\pi$, respectively, for the Pt JJ-based, Ta JJ-based, W JJ-based and Pd JJ-based SQUIDs. By comparing and correlating these measured (non-volatile) $\varphi_0$ with the zero-field Josephson supercurrent non-reciprocity (Figs. 2 and S5), we provide strong evidence for a direct link between these in our exchange-spin-split JJs. In particular, the estimated large values of $\varphi_0 = +0.44 \pm 0.02\pi$ ($+0.29 \pm 0.01\pi$) and $-0.33 \pm 0.02\pi$, respectively, for the Pt (Pd) and Ta JJs can account for their notably large $Q_{\mu_0 H = 0}$ with the opposite diode polarity. As shown in Fig. 5, the empirical linear scaling between the measured $\varphi_0$ and the obtained $Q_{\mu_0 H = 0}$ provides strong evidence that $\varphi_0$ is indeed the dominant factor contributing to $Q_{\mu_0 H = 0}$. According to recent theories (*36,37*), the low-$T$ limit ($T < 0.1T_c$) supercurrent diode efficiency of *intrinsic origin* can reach 20−30% at optimal magnetic fields, spin-orbit coupling and junction geometry. In this regard, we find that the *zero-field* diode efficiency of ≥ 15% obtained in our exchange-spin-split $\varphi_0$-junctions at 2 K ($T \sim 0.5T_c$) is promising.

We finally compare our results with a previous observation of the non-volatile $\varphi_0$ in SQUIDs made with symmetric Al/InAs nanowire/Al Josephson junctions (JJs) (20). In that case, unpaired spins arising from surface oxides or defects act as ferromagnetic impurities, giving rise to $\varphi_0$ non-volatility. The earlier nanowire $\varphi_0$ experiments thus rely on localized impurity spins interacting with the superconducting phase, which is fundamentally different from our approach that leverages engineered proximity magnetization combined with Rashba SOC. Although the observed $\varphi_0$ values are of the order of π, similar to ours, the weak controllability of surface spin density, spatial distribution, and spin orientation in the nanowire structure limits its practical prospects for large-scale integration. In contrast, while achieving reproducible interfacial magnetic ordering and controlled Rashba fields across wafers remains challenging and requires careful materials optimization, our work demonstrates that using conventional spintronic and superconducting materials with industry-compatible sputter deposition provides a higher degree of tunability and compatibility with wafer-scale fabrication.

## CONCLUSIONS

On the basis of exchange-spin-split JJs (*24-26*) and superconducting quantum interferometry (*12,17-20*), we have experimentally identified the proximity role of the Josephson $\varphi_0$-barrier in determining the zero-field supercurrent diode properties and demonstrated the presence of $\varphi_0$ (*20-22*) without $\mu_0 H$ and its direct link with the zero-field diode effect. We believe that our interferometry characterization of the non-volatile $\varphi_0$ provide a timely and important step for confirming the intrinsic $\varphi_0$-origin of the non-reciprocal Josephson supercurrents, for the $\varphi_0$-tuning of the zero-field diode performance (*i.e.* sign and magnitude), and for devising a new generation of the cryogenic non-volatile $\varphi_0$-memory. We expect that our results will provide a guideline for further technical improvements through the right choice of the proximity layer for the Josephson $\varphi_0$-barrier, as also pointed out in recent theories (*36,37*). Given the recent adoption of vortex-driven supercurrent rectifiers (*38,39*) in AC-to-DC converters with active lateral dimensions on the order of several micrometers for logic circuit applications, our $\varphi_0$-tunable Josephson diodes can enable the realization of ultra-low power, high-density circuitry, as the operating principle of our devices is free from the geometric constraints (*38,39*) required to accommodate Josephson or Abrikosov vortices and to induce their rectifying motion.

## METHODS AND EXPERIMENTAL

### Sample growth and device fabrication

Four different types of proximity layers of Pt(4−15 nm), Ta(4−15 nm), W(4−15 nm), and Pd(4−15 nm) were first sputter-grown on single-crystalline YIG(~200 nm) films (*5*) (Fig. S1) at room temperature by d.c. magnetron plasma sputtering in an ultra-high vacuum system with a base pressure of $1 \times 10^{-9}$ Torr. All these films were sputtered at 27 °C with a sputter power of more or less 15 W and at an Ar pressure of 3 mTorr, and were capped with a naturally oxidized $AlO_x$(1 nm) layer that was formed by sputter deposition of Al to prevent oxidation of the proximity layer.

To fabricate the lateral JJ and SQUID (Figs. 1A, 1B, and 1C), a central track of the proximity layer with lateral dimensions of $1.5 \times 50$ μm$^2$ is first defined using optical lithography, $Ar^+$-ion beam etching. We then defined electrical leads and bonding pads, formed from Au(60−80 nm)/Ru(2 nm), which were deposited by $Ar^+$-ion beam sputtering. Multiple Nb electrodes to form the lateral JJ and SQUID were subsequently defined on top of the central track via electron-beam lithography and lift-off steps. The Nb(50−60 nm) electrodes were grown by $Ar^+$-ion beam sputtering at an Ar pressure of $1.5 \times 10^{-4}$ mbar. Before sputtering the Nb electrodes, the $AlO_x$ capping layer and Au surface were Ar-ion beam etched away to make direct metallic electrical contacts. All lift-off processes were done with acetone in ultrasonic baths. We note that the edge-to-edge spacing $L$ between the adjacent Nb electrodes of the lateral JJ (Figs. 2 and 3) and SQUID (Figs. 4 and S5) was fixed at 80−100 nm to make the Josephson non-reciprocal supercurrents and anomalous phase-shift ($Q \propto \varphi_0 \propto L^2$, see Supplementary Material for details) detectably large at 2 K in a $^4$He cryostat.

**Measurements and analysis of Josephson *I*-*V* curves and SQUID data**

We measured current-voltage $I-V$ curves of the fabricated JJs and SQUIDs (Figs. 2, 3, and S4) and voltage-flux $V-\Phi_{SQUID}/\Phi_0$ curves of the $I$-biased SQUIDs with a 4-probe configuration in a Quantum Design Physical Property Measurement System (PPMS) using a Keithley 6221 current source and a Keithley 2182A nanovoltmeter. We carried out the demagnetization process of an unintentionally trapped magnetic flux in superconducting coils of the PPMS for > 3 hrs at $T$ > 10 K in advance to the measurements.

For the *zero-field* $I-V$ curve measurements (Figs. 2 and S5) below the junctions' $T_c$ (< 6 K), $\mu_0 H_{\parallel} = \pm$ 30 mT, much larger than the coercive field of YIG (*5*), is first applied along the *x*-axis and then returned to zero to preconfigure the remanent-state $M_{\varphi_0-barrier}$. The Josephson critical current $I_c$ was determined by fitting the measured $I-V$ curves with the standard formula for overdamped junctions (*34*), $V(I) = \frac{I}{|I|} R_n \sqrt{I^2 - I_c^2}$. When thermal noise/rounding effects $\left(\propto \frac{T}{I_c}\right)$ (*34*) on the $I-V$ curves become significant, especially for $\mu_0 H_{\parallel} \geq$ 30 mT or $T \geq$ 3.5 K, we determined the $I_c$ value at the point where $V(I) \approx$ 1 μV. We obtained the field-strength and field-angle dependences of $I_c(\mu_0 H_{\parallel})$ and $I_c(\gamma)$ (Fig. 3) by repeating the $I-V$ curve measurements at $T$ = 2 K at the applied $\mu_0 H_{\parallel}$, parallel to the interface plane of Nb electrodes. For a proximity layer with a fixed thickness of 4 nm, we conducted the *I-V* measurements on at least five different JJs on the same YIG/GGG(111) substrate. We found that over 80% of the working devices exhibited the same sign and a similar magnitude (±10%) of zero-field Josephson diode efficiency.

We measured the $V_{\pm}(\mu_0 H_{\perp}, I)$ curves for the $I-$biased SQUID by sweeping $\mu_0 H_{\perp}$ up and down up to $|\mu_0 H_{\perp}|$ = 4 mT. Note that as shown in our previous experiment (*35*) with a Cu JJ-based SQUID, the field drift error was found quite small (~0.01 mT). In Fig. 4, the measured $V_{\pm}(\Phi_{SQUID}/\Phi_0)$ data appear somewhat noisy. This noise is likely due to the challenge of precisely controlling the magnetic field at ~0.01 mT using a superconducting electromagnet in the PPMS, rather than issues with the amplitude of the voltage signals. Note also that given that the typical coercive field of YIG(~200 nm) fims at 2 K is on the order of 0.1 mT, and considering that the patterned proximity-magnetized Pt layer generally exhibits a larger coercive field than YIG, we can rule out the presence of unintentionally trapped flux (on the order of 0.01 mT) at $\mu_0 H_{\parallel}$ = 0 mT as a potential cause for the zero-field supercurrent diode effect and its polarity reversal in our system.

## ASSOCIATED CONTENT

### Supporting Information

Supporting Information is available free of charge at https://acs.nano/doi/xx.xxxx/. The Supporting Information includes an analytical expression for the diode efficiency in the limit of a small second harmonic and small $\varphi_0$; a quasiclassical theory of diffusive superconducting $\varphi_0$ Josephson junctions with a proximized Rashba normal metal and a spin-active interface with a magnetic insulator; and quantitative estimates of loop inductance and flux offset.

### Preprint

A preprint version of this work is available: Jeon, K.-R.; Kim, J.-K.; Yoon, J.; Jeon, J.-C.; Han, H.; Cottet, A.; Kontos, T.; Parkin, S. S. P. Proximity Engineering and Interferometric Quantification of a Non-Volatile Anomalous Phase Shift in Zero-Field Polarity-Reversible Josephson Diodes. arXiv 2025, arXiv:2505.20598. https://arxiv.org/abs/2505.20598 (accessed January 8, 2026).

## ACKNOWLEDGMENTS

**Funding:** This work was supported by the Max Planck Partner Group 2023, the POSCO Science Fellowship 2024, the National Research Foundation (NRF) of Korea (Grant No. 2020R1A5A1016518) and the European Union (ERC Advanced Grant SUPERMINT, project number 101054860). **Author contributions:** K.-R.J. and S.S.P.P conceived and designed the experiments. K.-R.J. fabricated the lateral Josephson junctions, Hall-bar and non-local magnon devices with help from J.-K.K., J.Y. and J.-C.J., and carried out the transport measurements with help of J.-K.K. and J.-C.J. H.H. contributed to the structural analysis. K.-R.J. performed the data analysis while A.C. and T.K. developed a comprehensive theory for the exchange-spin-split Rashba-type $\varphi_0$-junction. K.-R.J., A.C., T.K. and S.S.P.P. wrote the manuscript with input from all the other co-authors. **Competing interests:** The authors declare no competing interests. **Data and materials availability:** All data needed to evaluate the conclusions in the paper are present in the paper and/or the Supplementary Materials.

## FIGURE CAPTIONS

**Fig. 1. Josephson $\varphi_0$-junction and SQUID fabricated on the same proximity layer.** (**A**) Schematic illustration of the Josephson $\varphi_0$-junction and measurement configuration. Here $L$ is the length of the Josephson $\varphi_0$-barrier. Notice that $\gamma$ is the azimuthal angle of the magnetization of the $\varphi_0$-barrier (// $\mu_0 H_{\parallel}$ = IP magnetic field) in the clockwise direction with respect to the *d.c.* bias-current $I$ along the $+y$ axis. (**B**) Schematic of the superconducting quantum interferometer device (SQUID) and measurement scheme. Here $L_{x/y}^{ctc}$ is the centre-to-centre spacing between the tracks defining the two opposite sides of the SQUID. (**C**) Scanning electron micrograph of the fabricated Josephson $\varphi_0$-junction and SQUID on top of the same proximity layer, which is proximity-magnetized by a ferrimagnetic insulator $Y_3Fe_5O_{12}$ (YIG) underneath it. Note that in the SQUID geometry, supercurrent flow directions ($I_s^1$, $I_s^2$) of the constituent Josephson $\varphi_0$-junctions are orthogonal to each other, allowing for measurements of both non-volatile $\varphi_0$ and zero-field supercurrent diode efficiency $Q_{\mu_0 H = 0}$ in a single SQUID [12,20]. In (C), the scale bar is 1 μm.

**Fig. 2. Zero-field polarity-switchable Josephson diodes with a different proximity layer.** (**A**) Zero-field current-voltage *I-V* curves of the magnetic 4-nm-thick Pt Josephson junction (JJ) for two different Pt magnetization $M_{Pt}$ (// $\pm x$-axis) below the junction's superconducting

transition temperature $T_c$. Here $I$ is normalized by $I_c^{avg}$ $\left(= \frac{I_c^+ + |I_c^-|}{2}\right)$ = 70−105 μA. In the yellow (cyan) shaded regime, the Josephson supercurrent flows only in the positive (negative) *y*-direction, as indicated by the diode symbols. Notice that the displayed *I-V* curves were obtained by sweeping *I* forward and backward, *e.g.,* 0 mA -> +0.15 mA -> −0.15 mA -> 0 mA. (**B**) and (**C**) Data equivalent to (A) but for the JJ with a different 5d HM barrier. In (B) [(C)], the 4-nm-thick Ta (W) proximity layer is employed. Note that the zero-field diode polarity of Ta and W JJs is found to be opposite to that of Pt JJ. (**D**) Data equivalent to (A) but for the JJ with a 4d Pd(4 nm) proximity layer with high (para-)magnetic susceptibility (*30,31*). (**E**) Zero-field diode efficiency $Q_{\mu_0 H = 0} = \frac{\Delta I_c}{2 I_c^{avg}}\Big|_{\mu_0 H = 0}$ as a function of normalized temperature $T/T_c$ for the Pt, Ta, W and Pd JJs.

**Fig. 3. Magnetic field-strength and angle dependences of the Josephson diode efficiency.** (**A**) Josephson diode efficiency $Q = \frac{\Delta I_c}{2 I_c^{avg}} \left(= \frac{I_c^+ - |I_c^-|}{I_c^+ + |I_c^-|}\right)$ versus in-plane (IP) magnetic-field strength $\mu_0 H_{\parallel}$ plot for the magnetic 4-nm-thick Pt Josephson junction (JJ), taken at the fixed azimuthal angle $\gamma = 90^0$ of the Pt magnetization $M_{Pt}$ with respect to the positive *y*-axis. Notice that the orange and cyan symbols represent, respectively, the sweep-up and sweep-down directions of $\mu_0 H_{\parallel}$ and the error bars for the applied $\mu_0 H_{\parallel}$ (approximately 0.01 mT) are much smaller than the size of the data points. The inset shows a magnified plot around $\mu_0 H_{\parallel} = 0$. (**B**) and (**C**) Data equivalent to (A) but for the JJ with a different 5d HM layer. In (B) [(C)], the 4-nm-thick Ta (W) proximity layer is used. (**D**) Data equivalent to (A) but for the JJ with a 4d Pd(4 nm) proximity layer with high (para-)magnetic susceptibility (*30,31*). (**E**) $\gamma$-dependent

$\frac{\Delta I_c}{2I_c^{avg}}$ of the Pt, Ta, W and Pd JJs at the constant $\mu_0 H_\parallel$ = 5 mT, which is applied to rotate $M_{Pt}$ (// $\mu_0 H_\parallel$) in the plane. Since the strength of exchange-spin-splitting fields in our proximity system is greater than $\mu_0 H_\parallel$ = 5 mT, $\varphi_0$ can be assumed to be constant for this measurement.

**Fig. 4. Probing $\boldsymbol{\varphi_0}$ by SQUIDs and its correlation with the zero-field diode efficiency.** (**A**) Time-averaged voltage $V_\pm$ as a function of perpendicular magnetic field $\mu_0 H_\perp$ for the d.c. current *I*-biased SQUID, which consists of Pt(4 nm) JJs. Here the applied d.c. bias *I* is in the range of 145−210 μA. (**B**) Low-field *V* oscillation as a function of normalized magnetic flux $\Phi_{SQUID}/\Phi_0$, corresponding to the wine shaded regime in (A). (**C**) and (**D**) Data equivalent, respectively, to (A) and (B) but for the SQUID with Ta(4 nm) JJs. (**E**) and (**F**) Data equivalent, respectively, to (A) and (B) but for the SQUID with W(4 nm) JJs. (**G**) and (**H**) Data equivalent, respectively, to (A) and (B) but for the SQUID with Pd(4 nm) JJs. Notice that for the Pd JJ-based SQUID, a non-negligible parabolic-like background signal is subtracted from (G) to better estimate the non-volatile $\varphi_0$ of our magnetic Pd JJs in (H).

**Fig 5. Proportionality of $\boldsymbol{\varphi_0}$ to the zero-field diode efficiency $\boldsymbol{Q_{\mu_0 H = 0}}$.** The obtained $Q_{\mu_0 H = 0}$ (Fig. 2) is plotted as a function of the measured $\varphi_0$ (Fig. 4), revealing a clear empirical linear relationship. Note that the error bars for $\varphi_0$ and $Q_{\mu_0 H = 0}$ are smaller than the size of the data points.

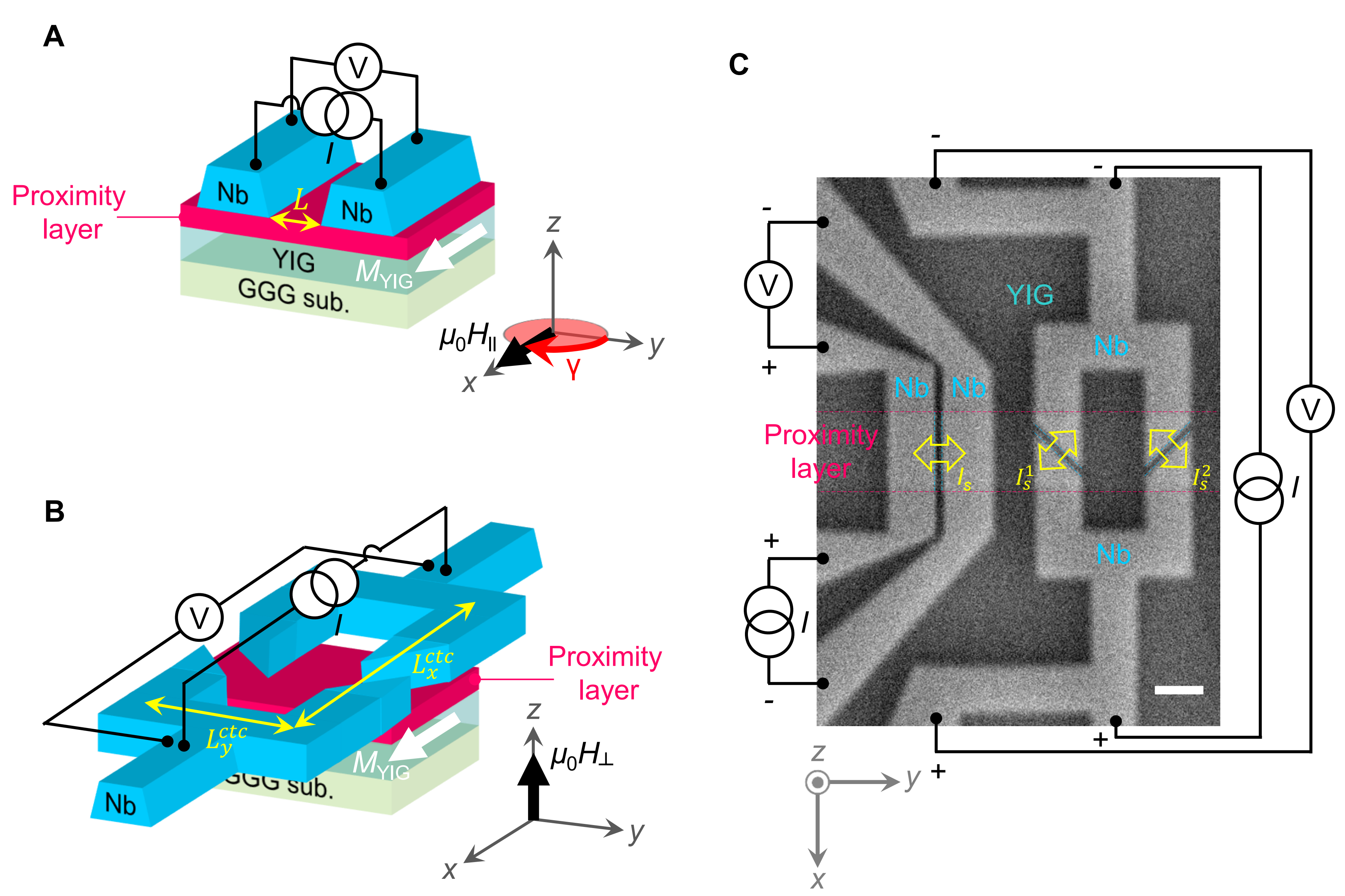

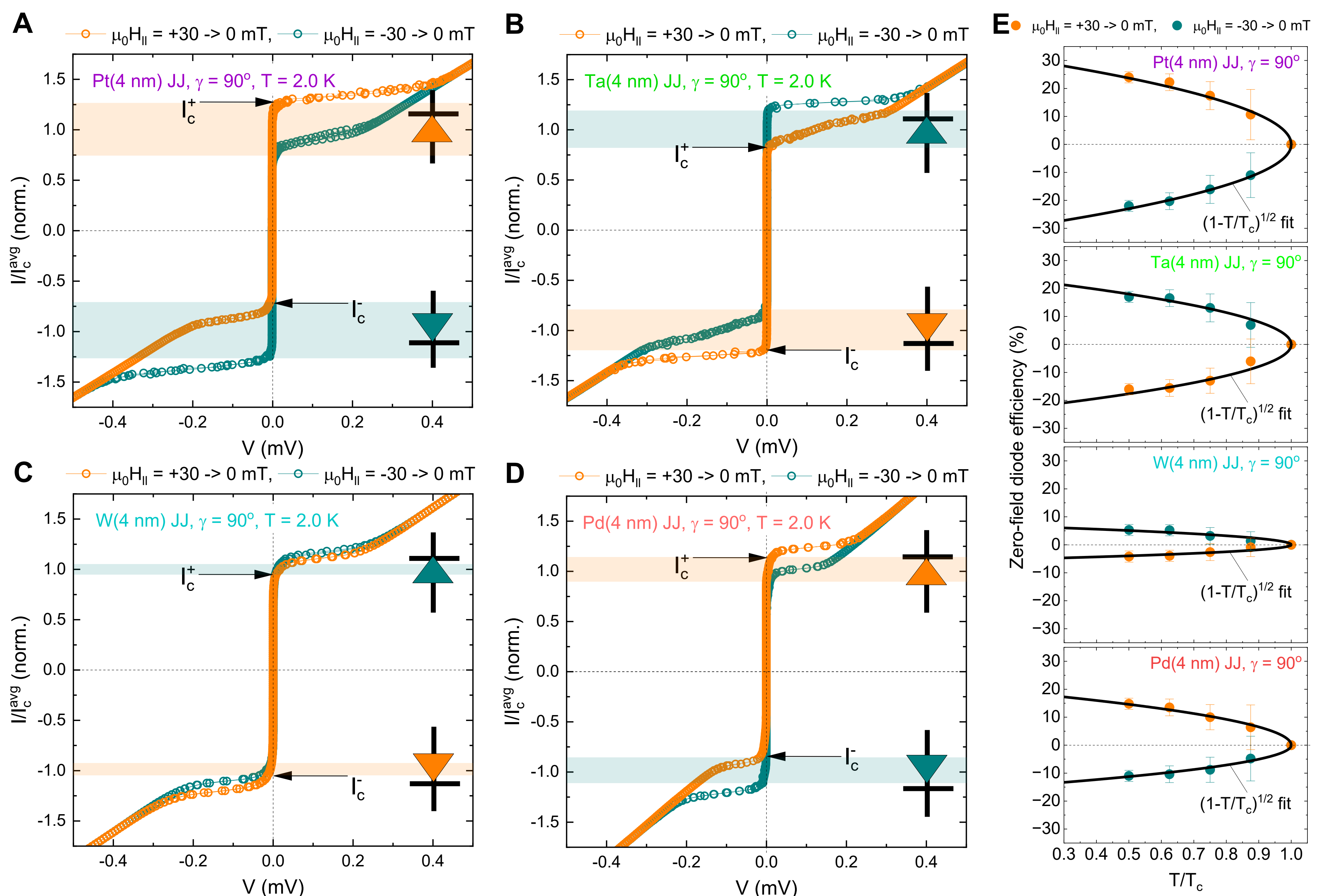

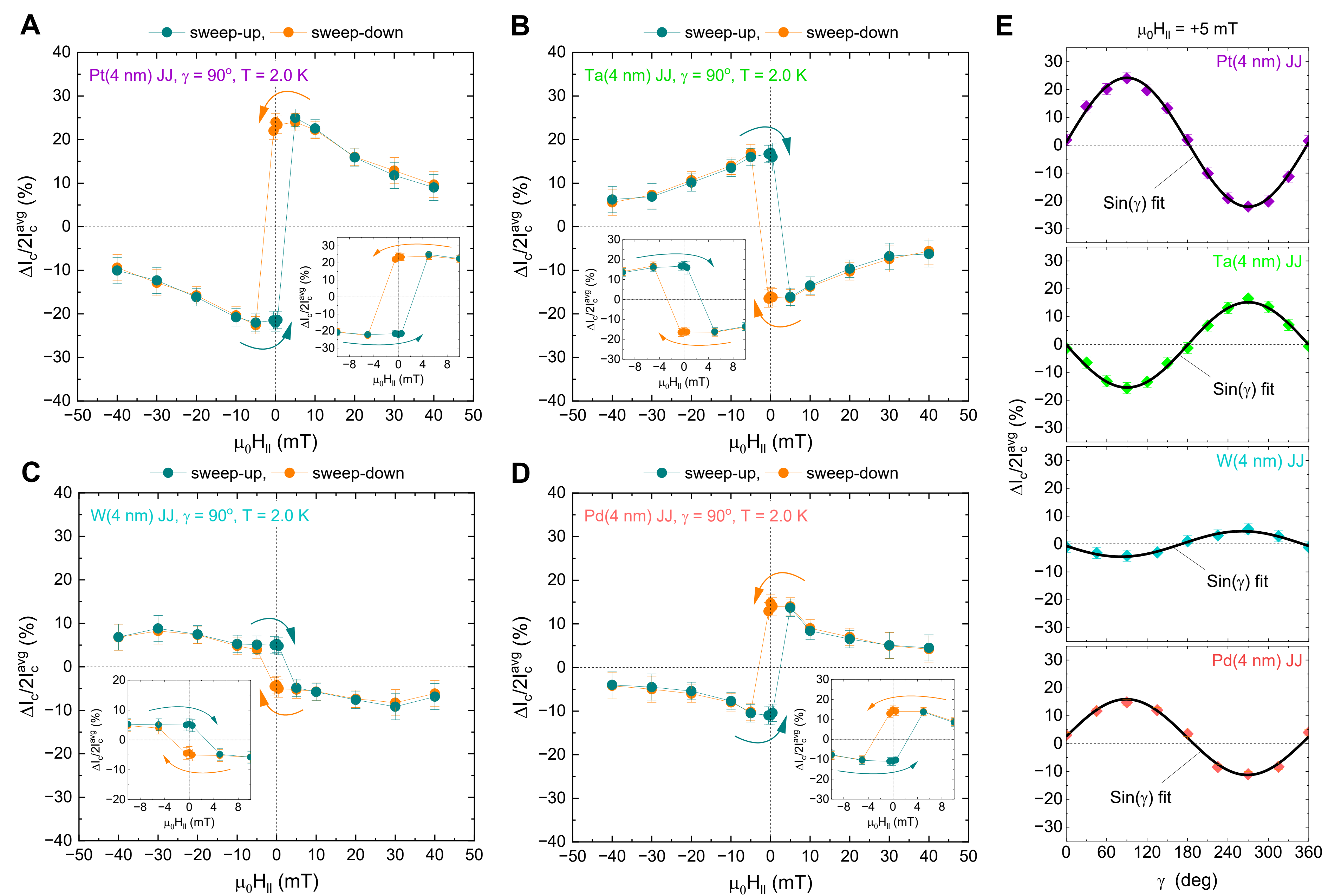

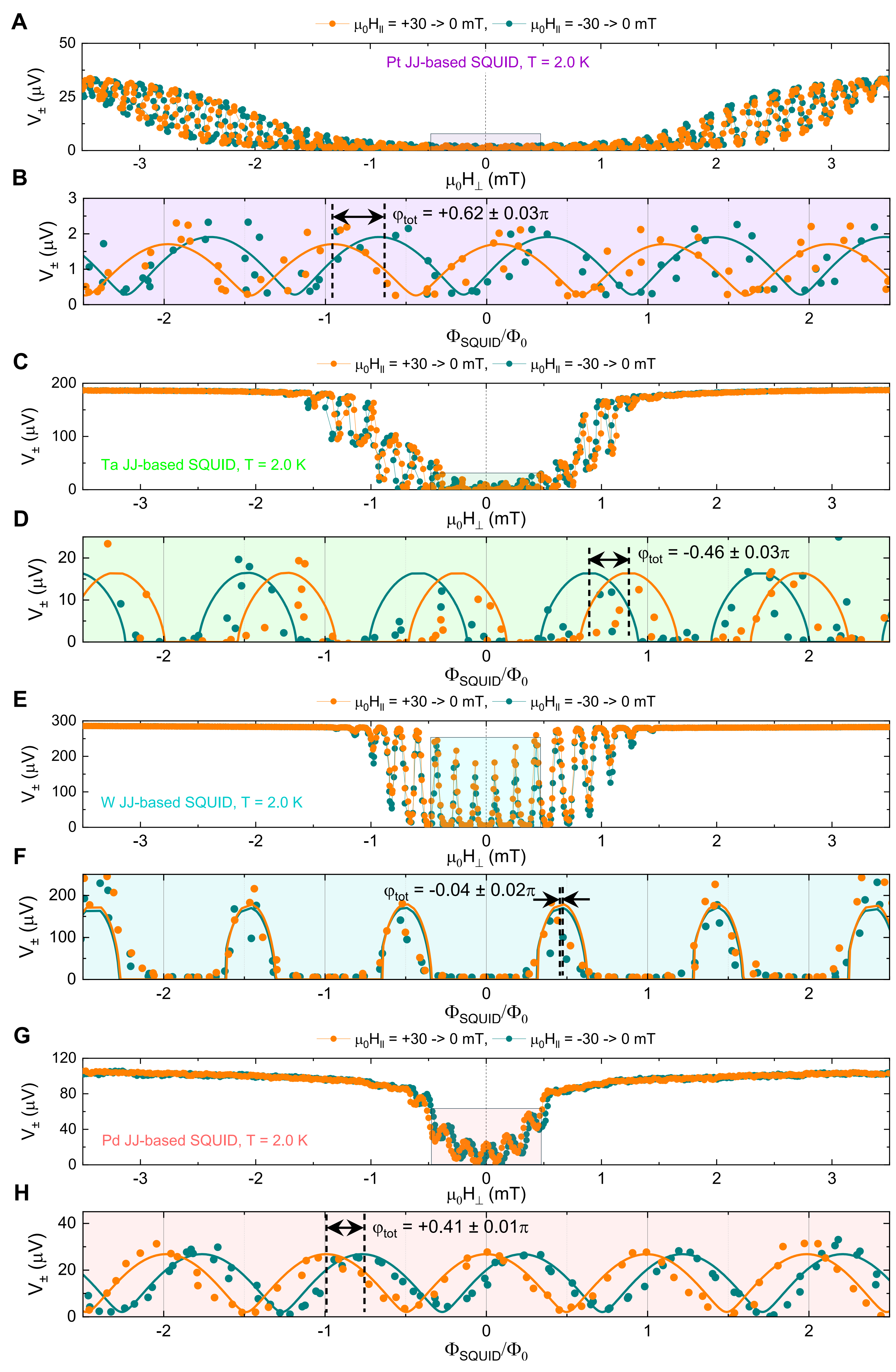

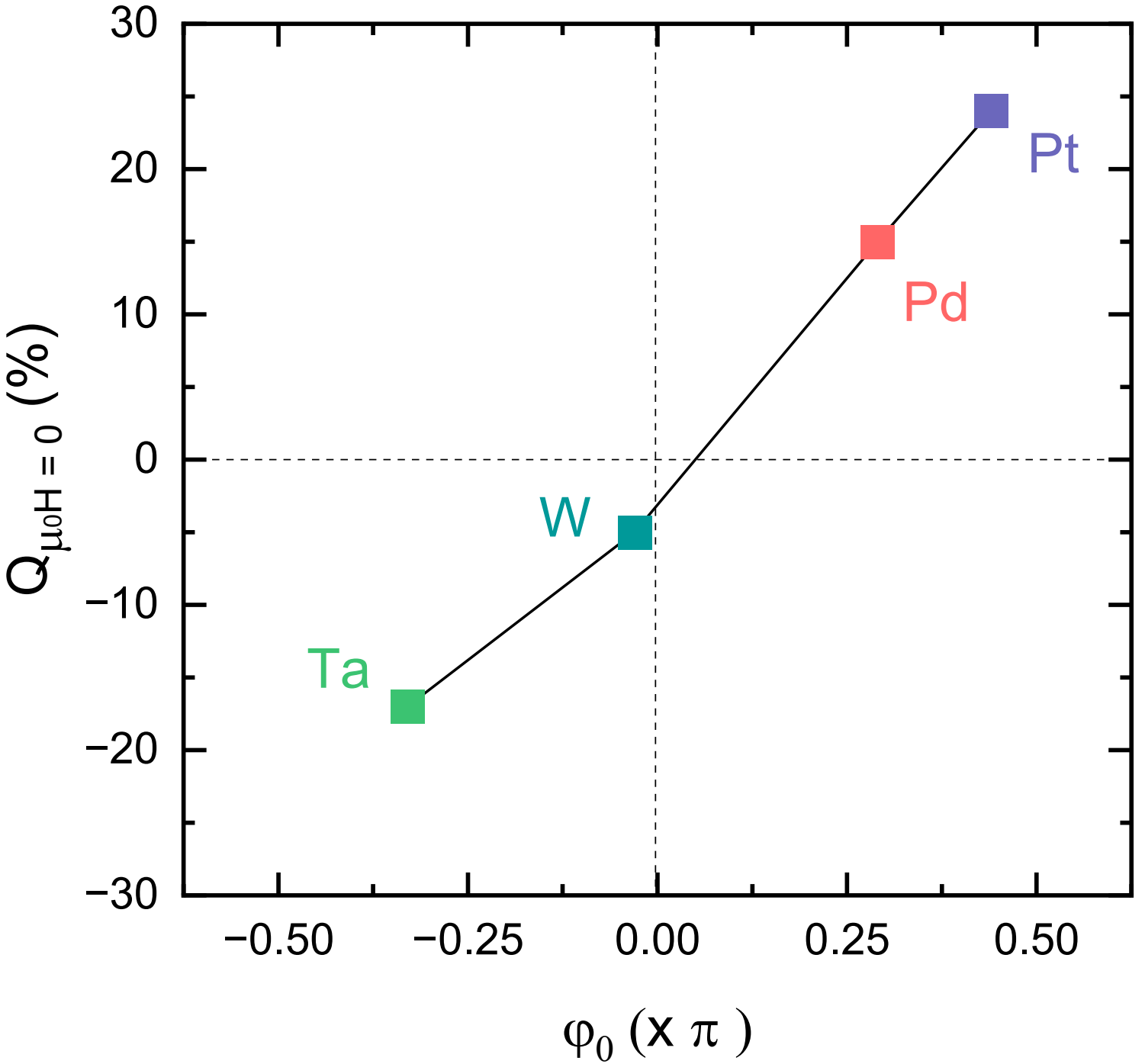